\documentclass[10pt, conference, compsocconf]{IEEEtran}

\usepackage{graphics}
\usepackage{graphicx}
\usepackage{epsfig}
\usepackage{amssymb}
\usepackage{lineno}
\usepackage[hyphens]{url}

\usepackage{changes}
\definechangesauthor[color=red]{is}
\definechangesauthor[color=blue]{sd}
\definechangesauthor[color=green]{rs}

\newcommand{\NN}{\mathcal{N}}
\newcommand{\RR}{\mathcal{R}}

\newcommand{\CC}{\mathcal{C}}
\newcommand{\SCL}{\mathcal{S}}
\newcommand{\TT}{\mathcal{T}}
\newcommand{\PP}{\mathcal{P}}
\newcommand{\LL}{\mathcal{L}}

\usepackage{float}
\title{Spatio-temporal prediction of crimes using network analytic approach}

\author{
    \IEEEauthorblockN{Saroj Kumar Dash\IEEEauthorrefmark{1}\IEEEauthorrefmark{3},
    Ilya Safro\IEEEauthorrefmark{1}\IEEEauthorrefmark{5},
    Ravisutha Sakrepatna Srinivasamurthy\IEEEauthorrefmark{2}\IEEEauthorrefmark{4}}
    \IEEEauthorblockA{\IEEEauthorrefmark{1}School of Computing, Clemson University}
    \IEEEauthorblockA{\IEEEauthorrefmark{2}Department of Electrical and Computer Engineering, Clemson University\\ Email: \{ \IEEEauthorrefmark{3}sdash, \IEEEauthorrefmark{5}isafro, \IEEEauthorrefmark{4}rsakrep \}@clemson.edu}    
}

\begin{document}


\maketitle

\begin{abstract}
It is quite evident that majority of the population lives in urban area today than in any time of the human history. This trend seems to increase in coming years. A study \cite{urban} says that nearly  80.7\% of total population in USA stays in urban area. By 2030 nearly 60\% of the population in the world will live in or move to cities. With the increase in urban population, it is important to keep an eye on criminal activities. By doing so, governments can enforce intelligent policing systems and hence many government agencies and local authorities have made the crime data publicly available. In this paper, we analyze Chicago city crime data fused with other social information sources using network analytic techniques to predict criminal activity for the next year. We observe that as we add more layers of data which represent different aspects of the society, the quality of prediction is improved. Our prediction models not just predict total number of crimes for the whole Chicago city, rather they predict number of  crimes for all types of crimes and for different regions in City of Chicago.\\
Reproducibility: our code is available at \url{https://goo.gl/V1yVLk}.\\ 

\end{abstract}







\section{Introduction}
It is quite evident that majority of the population lives in cities \cite{urban}. 
Modern and effective police system plays a key role in city safety that is especially critical in megalopolises with culturally diverse population.
In the era of big data, analyzing historical events of  criminal activity might give some critical information to implement intelligent policing across the city. Many cities have realized this and crime data have been made publicly available for research. 
There exist different types and formats of crime datasets \cite{dataset,dataset1,dataset2} which, however, can be summarized as collections of crime records that typically contain time, location and short description type of the crimes without providing too many details due to such obvious reasons as privacy and confidentiality. We discuss some of them in Section \ref{related_works}.
However, not much research has been done to improve the crime prediction by fusing the crime data with other types of city data which is the goal of our research. 

\subsection*{Our contribution} In this paper, we analyze the Chicago crime dataset and show how the quality of crime prediction can be improved if fused with other city datasets \cite{dataset}. 
In contrast to previous approaches, we use network analytic techniques to connect different regions of the city and fuse them with different data types to predict next crime pattern in a given region. We fuse crime data with such information as sanitation, schools, police stations, libraries, and 311 services to make train the prediction models. We observe that using different types of data which are not directly related to crime increases the accuracy of models and information from distant regions of the city may give insight on criminal activities of another distant part of the city.

This paper is organized as follows. We discuss the related work in which relevant spatial and temporal predictive modeling is discussed in Section \ref{related_works}.
We introduce our prediction method including necessary definitions and datasets used in Section \ref{experiment}. The computational results that cover the entire Chicago area and all types of crimes are presented in Section \ref{results}. Finally, we conclude and discuss potential research directions in Section \ref{future}.

\section{Related Work}\label{related_works}
The area of crime data mining has shown increasing popularity and demand in recent years. One of the most important reasons is the increasing crime rates. 
The ``Economist'' indicates that there was a 22\% increase in the number of murders in Chicago in recent times \cite{increase}. 
As a result of massive urbanization, the situation with crime in many cities is similar to that in Chicago. Consequently, there was a rapid increase in research on predictive crime analysis for better policing and to stop gang activities or any other major criminal activities. To predict crime, several classical statistical techniques are being applied, whose results are then used to take correct steps taken by the police to prevent crimes.   
Statistical methods such as the auto-regression modeling \cite{role}, association of crime events \cite{data1}, and random walks \cite{rwalk} are being used in state of the art analysis of crimes. In this section, we review several relevant  crime analysis methods.
   
   In \cite{data1}, Brown and Hagen discuss an associative approach to find similarity among the crime records in the available data. The authors find rules to  associate each crime event data with importance attributes which qualitatively and  quantitatively adds weights to the attributes. The total similarity measure between crime records is computed which is the summation of the products of the weights of that attribute and the normal association value between the two crime event attributes. The weights are calculated based on the importance of the attribute in a particular crime record that is considered for finding the similarity. 
Once the weights are calculated, the \emph{total similarity measure} (TSM)  is computed to find the crimes which are similar to each other in nature. For an interpretability of crime attributes, a similarity measure of attributes was developed. 
Then using the Bayers Rule method, a relative association between different crime records is found. This method is compared to the manual crime similarity search which is being used in the police department so far. The method was found useful for its improved performance and applicability in the crime trend analysis.

   In \cite{rwalk}, the goal of Tayebi et al. is to predict the probability of an activity space of a criminal. Similar to our approach, the authors  use network analysis to predict the crime activity area at certain location. The paper is based upon hotspot (high intensity crime area) recognition of the criminal and then the spatial information for building an activity space of the criminal. This spatial criminal activity space is used to see the activity of the criminal and then for taking appropriate actions. It is based upon the basic thinking that a criminal does more crime in a vicinity where the criminal is more familiar with and also where the victim is new. Based on the hotspot recognition and pattern, the next crime location is predicted. It uses the random walk model to predict the criminals home location and based upon that, the method identifies the next crime location. This method is also mentioned in all the network analysis approaches discussed in \cite{police1}.

Spatial hot spot analysis in one of the main streams in crime prediction given a historic data for more than two decades. The idea introduced in \cite{block1995stac} can be summarized as clustering training criminal data into hot spots using mixture models and density estimation with the hypothesis that crimes are likely to happen within such clusters (hot spots). However, these methods do not take into account socio-environmental information such community quality, people awareness, police station proximity, and quality of schools. Also, lack of historical information on particular areas make such models useless on them. Other works in this class of methods include \cite{liu2003criminal,smith2007discrete}. 
Our work is complementary to such methods because, using the network analytic approach, we find similar areas whose historical information is also used in prediction.
 
    The objective of \cite{role} is similar to ours. The authors take the Chicago crime dataset (which is also part of our input) and creates the time series data to predict number of crimes at a \emph{specific} area on a week by week basis. The experiment is carried out in a particular Eastern region of Chicago. All of the crime data from 2001 to 2012 is used to train the model which is used to predict the crime numbers for year 2013. For this goal, a statistical method named  \emph{autoregressive integrated moving average} (ARIMA) was applied. This method differentiates itself from the normal regressive method by subtracting a differencing transformation to the data before applying the regression. The error subtraction in ARIMA makes the prediction more unbiased as the errors are always being subtracted from the data.  The prediction has an error rate of 14.7\%. For a one-year of prediction of the number of crimes, it showed an accuracy of 84\%. For the second year, the accuracy was dropped to 80\%. We also use the autoregression method as one among the three methods to verify the quality of the crime  prediction on our fused data.


In \cite{embedded}, Bastomski et al. consider structural embeddedness within co-offending network to predict homicide rate. According to this paper, the social structure of a neighborhood typically impacts the number of crimes that occur in that neighborhood. If two neighborhoods are structurally similar (similar economic conditions, similar quality of education, etc.) then the types of crime and number of crimes will also be similar. \emph{We use and extend this observation in our methods in which we assume that a prediction of crimes can be made using similarities of several attributes of neighborhoods.} In particular, it is important to mention that these neighborhoods need not be adjacent or connected spatially. This also indicates a better chance of crime spread between similar neighborhoods. The network formed when two or more individuals engage in a criminal activity is known as co-offending network. This paper considers residency of co-offenders as nodes and the neighborhoods belonging to these co-offenders as edges. Using this network, the $k$-core measure is used as an independent parameter and with the help of other measures, the parameter for embeddedness 
is formed. Then by using ordinary least squares, 
the homicides rates are predicted. Finally, the paper confirms that inter-neighborhood ties are not explained by geographic proximity alone.
    
An approach to predict links between pairs of individuals with examples in the Italian mafia was proposed by Bersluconi et al. in \cite{linkprediction}. In this paper, the individuals are considered to be the nodes and they are connected based on the number of telephone calls. Three networks are considered, namely,  \textit{wiretap records network}, \textit{arrest warrant network} and \textit{judgment networks}. Weight of the links in the wiretap records network is proportional to the number of phone calls between the individuals. Arrest warrant network is obtained by reducing  wiretap records. Similarly, judgment network is obtained by reducing arrest warrant network. This reduction is done by removing the connection between individuals if the number of calls is small. The authors argue that the removal of links is characterized by low betweenness in the network. That is, even if a link is removed, the corresponding  individuals are connected indirectly in some way. The second hypothesis is that nodes are likely to be connected if they are similar. In this case, individuals who collaborate rarely will have dissimilar attributes such as interests and background. This hypothesis is proved to be correct and hence author uses this concept to predict missing links. If two nodes are have high similarity scores, there is a chance of connection between these criminals. Author also tests Katz index similarity and structural perturbation method which also gives similar results.
    
In \cite{temporal}, Almanie et al. develop a method to predict criminal hot spots and prominent crime types in a given area and also the frequency of crimes for a given time period. 

This paper applies three methods, namely, the Apriori algorithm, Naive Bayes and Decision Tree classifiers on the Denver and Los Angeles crime datasets. They consider location, time, crime type data to predict the frequency of crime events. Out of all methods applied, the results of Naive Bayes were found of greatest accuracy which is 54\%. The paper gives a basis  to apply statistical methods to the crime data and also to consider the spatial data for crime prediction.

    
 

In \cite{predict_crime}, Henderson et al. propose a very unique method in which  market prediction algorithm \cite{marketprediction} is applied to predict Chicago crime rates. The authors hypothesize that if a crime market is setup where people can trade contracts on crime rates, the market with the help of information gathered from people can be used to predict the crime rates. Market prediction employs a natural feedback loop. If someone believes (on the basis of some information) that crime rate may go up, then she will bet highly on the increase in crime rate and vice versa. Hence, it forms a natural feedback loop. 

\section{Crime prediction method}\label{experiment}
  
  Our objective is to predict crime levels for different types of crimes and a given year based on previous years criminal activity and other social aspects. We observe that augmenting the crime data with  other social aspects such as education, and economic conditions give few insights for predicting the crime and improve the quality of prediction. Also, social aspects of other similar communities give out critical information for predicting the crime pattern for next year. The experiments are conducted on the Chicago crime dataset \cite{dataset} provided by the Police Department for public use.
  
A critical component of the proposed method is to  fuse various types of social and historical information in a network which helps to find relationships between different communities within the city. The extracted relationships are incorporated in our prediction model. Model performance is demonstrated on predicting crime rates for year 2015 and compared with similar models that are not augmented with additional social information and relationships between communities. We also compare our results with previously proposed autoregressive model \cite{rwalk}  which does not take any information other than historical crime data into consideration. In the following subsections, we  discuss the datasets used, an approach to the problem, and the prediction model.
  
\subsection{Datasets}\label{sub:dataset}
	\par The City of Chicago provides the city data (including the crime dataset) for public use through the Chicago Data Portal at \url{http://data.cityofchicago.org/}. The crime dataset spans from a time period of 2001 to present and has record of all criminal activities happened in Chicago city except for last seven days as the data is updated every week. Below is the list of all the data sources that we have used in our analysis. 

\begin{itemize}
	\item Crime dataset includes 6.6 million crime data records from Chicago from 2001 to present. The total size of data is 1.41 GB.
	\item Chicago library data includes library locations and details with geographical coordinates of 80 libraries in Chicago city.
	\item Library visitors by location: This data gives information about the number of visitors in each month for all 80 libraries in Chicago city.
	\item Police station data: Gives us an information about the location of 25 police stations for 25 districts in Chicago city. 
	\item Police districts boundaries: The areas for which a police station is responsible for. Used to connect communities, crimes and police stations. 
	\item Dataset of all 311 service (non-emergency municipal services) requests. This includes the all-lights service requests, transit ride requests, pot holes requests, sanitation community requests and vacant house information. The size of this data is approximately 450 MB.
	\item Chicago public school progress information: This data is available from 2011 to 2017. This dataset contains locations and related data of 188 schools in the Chicago area. 
    \item Chicago school average ACT (American College Testing) scores: The average ACT scores were collected for all the 188 schools. 
\end{itemize} 

The crime data is available from year 2001 till now. However, most of the other data can be traced back to 2011 only, so for the training of our models, we used all available data from 2011 to 2014. The test sets were created using the 2015 data.

\subsection{Approach}\label{sub:approach}
\begin{figure}
\includegraphics[width=0.4\textwidth]{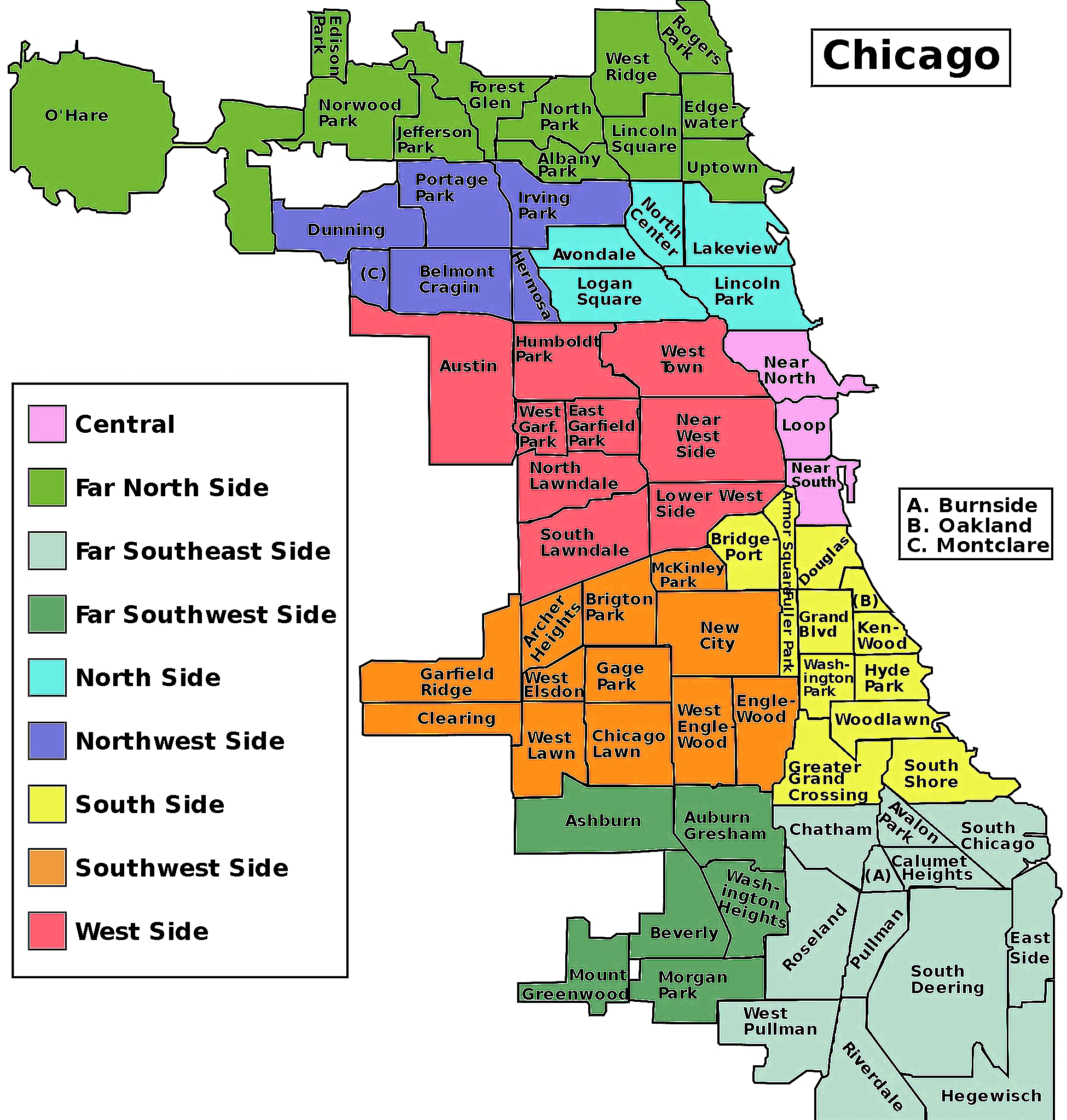}
\caption{Chicago is divided in 77 community areas which are well defined and do not overlap. Each community area has one or more neighborhoods in it. Image courtesy  www.thechicago77.com}\label{fig:chicago}
\end{figure}

The objective of this work was not just to predict the overall crime activity in Chicago city. Instead, we are interested in predicting the crime activity for \emph{various regions} of Chicago city. There are several ways to divide Chicago city into geographical regions and one of them, defined by the Social Science Research Committee at the University of Chicago, is called \emph{community division} \cite{ssrcAreas}. Chicago can be divided into 77 communities and the map of the communities in Chicago is readily available from City of Chicago portal (see Figure \ref{fig:chicago}). The communities are officially recognized by the City of Chicago. The 77 communities can be seen in the city of Chicago Portal for Dataset \cite{dataset}. The community number in the dataset tells about the  spatial resolution of Chicago. This spatial resolution could be improvedlater by considering the exact geospatial coordinates of the crime reported. For simplicity we have considered the resolution of the community area in our experiments. We found that among all the datasets that we mentioned in the above section the coordinates or the spatial location of the crime was always available in all the different datasets. Thus we took community or the location of crime reported as a unifying attribute to build the network by connecting all the datasets through there location. 
Because the reasons of illegal activities depend on social factors, it is clear that two communities might share similarity in crime patterns if they are neighbors to each other since social factors typically do not change drastically between neighboring communities.
However, it has been observed that two communities can be similar in crime activity even if they are not neighbors. This is because both communities can have similar social structure \cite{embedded} which may help with a predictability of community attributes (or features). To explore the social structure, we built a network of 77 communities and computed the similarity matrix between the communities which was used to extract two most similar communities for each reference community. All  predictions described below are computed using the information from two most similar communities.

\par 
Once we get features to predict criminal activity for the given year, we use two models to make the prediction. Those are polynomial regression and support vector machine. We compare our results with auto-regression model as it is just an extrapolation algorithm. The results will be discussed in section \ref{results}.

\subsubsection{Building Network}\label{sub:network}
We define a multi-layered network $\NN$ with an underlying edge-weighted undirected graph $G=(V,E)$ with $|V|=n$ and $|E|=m$,  in which we fuse different  information sources. The set of nodes is defined as a union of several layers
\[
V = \CC\cup \TT \cup \SCL \cup \PP \cup \LL \cup \RR,
\]
where, in our specific experiments with Chicago city, 
\begin{itemize}
\item $\CC$ is the set of 77 communities.
\item $\TT$ is the set of crime  primary types defined by the Police Department (represented by Illinois Uniform Crime Reporting (IUCR) Codes \cite{dataset}), $|\TT|=34$. Examples of the primary types include theft, narcotics and robbery. 
\item $\SCL$ is the set of 188 schools. 
\item $\PP$ is the set of 25 police stations. 
\item $\LL$ is the set of 80 libraries in Chicago.
\item $\RR$ is the set of 8 types of 311 service requests, namely, sanity, pot holes, lights one, light all out, lights alley, trees, and vacant.
\end{itemize}
\begin{figure}
\centering
\includegraphics[width=0.5\textwidth]{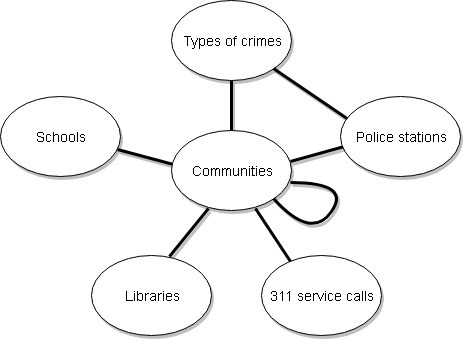}
\caption{Schematic representation of all node and edge types.}\label{fig:backbone}
\end{figure}
The set of edges $E$ (see schematic backbone of the network in Figure \ref{fig:backbone}) includes different types of connections between nodes, namely,
\begin{itemize}
\item  A community $c\in \CC$ is connected to a crime type $t\in \TT$ if that $t$ was registered in $c$ and number of such crimes determines the edge weight. Communities are also connected with each other if they share a border.
\item A school $s\in \SCL$ is connected to community $c\in \CC$ if $s$ is located in $c$. Edge weights between schools and communities are determined by the schools average ACT scores. The higher the ACT scores, the higher the standard of education.
\item Police stations can belong to more than one community. A police station $p\in \PP$ is connected to all communities in $\CC$ that it belongs to according to \cite{dataset} and also to the neighboring communities in case they are not connected directly. 

\item The police station node $p\in \PP$ is connected to crime type $t\in \TT$ based on where the crime has happened. The $pt$ edge weight equals the number of times crime type $t$ has been reported to police station $p$.

\item Each library $l \in \LL$ is connected to its community $c\in \CC$ in which it is located. The edge weight is the number of visitors visiting that library. Large edge weights indicate educated community and, hence, may impact levels and types of crime.
\item Every service type request to 311, $r \in \RR$, is connected to its community $c \in \CC$ in which request was registered. The edge weight is the number of requests of type $r$ in $c$. A big number of requests indicates that community is active in fixing their issues.
\end{itemize}

\begin{figure}
\centering
	\includegraphics[scale=0.25]{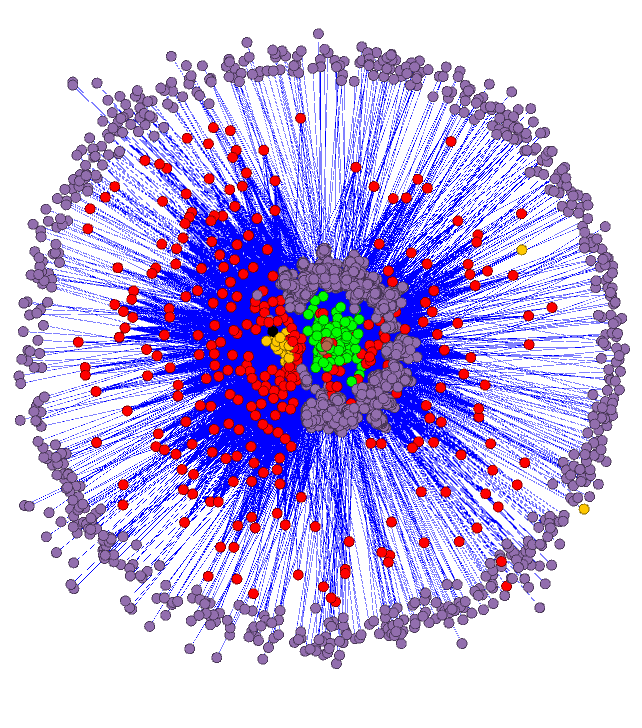}
	\caption{Visualization of the complete network for year 2011. Shows complete network with crime type nodes and the police network connected to each other.Nodes for this network are community(green), crime types(red), Schools(violet),Police stations(orange) and sanitations(brown).}
	\label{fig:fig1}
\end{figure}

The weighting function on edges is denoted by $w:E\rightarrow \mathbb{R}_{>0}$. Prior adding all layers of edges to $G$, they are normalized with respect to themselves using the min-max normalization. This normalization brings all values to $[0,1]$ interval and is important to avoid numerical problems that affect the prediction algorithms. An example of a resulting network is visualized in Figure \ref{fig:fig1}.

All events described in our datasets contain information about the time. 
For our analysis we build one network for each month for the period of 2011-2015 (total 60 networks, $\NN_i$, $1\leq i \leq 60$). The dynamic nature of criminal activities is addressed with the analysis of all $\NN_i$ from which corresponding similarity matrices $M_i$ (see Section \ref{sec:sim}) are built.

\subsubsection{Similarity}\label{sec:sim}
To get the top two similar communities for each community in $\CC$ and to generate similarity matrices $\NN_i$ per month, we used random walk similarity \cite{rwalk}. Random walk similarity is a metric frequently used in network analysis in which the Markov chain model is defined on transition probability for a random walk to move from one node to another. Defining a transition matrix $P\in \mathbb{R}^{n \times n}$, where $p_{ij} = \frac{w_{ij}}{\sum_k w_{ik}}$ is the probability of transition from $i$ to $j$, we can compute the similarity between two nodes in $V$ using the pseudo-inverse of the Laplacian  of $G$ \cite{pseudoinverse}.

Using the above method to compute the random walk similarity between two nodes we obtained the similarity between all pairs of communities and formed a similarity matrix. The features from top two similar communities are also used as features for the prediction model explained in the next section. It is important to mention that our experimental observation confirms that neighbor communities may not always be similar and similar communities may not always be neighbors. 

\subsection{Features}
Besides the crime data, for each community, we define additional features for the prediction models. These features include the number of police stations in that community, number of 311 service calls registered, number of schools and libraries that are connected to that community. We also consider this information for top two similar communities. Below we describe the set of features for each data point $p_{i,c}$ that corresponds observations in month $i$, $1\leq i \leq 60$, in community $c\in \CC$ whose top two similar communities are found to be $a,b\in \CC$. The vector $p_{i,c}$ is a concatenation of the following vectors and values:
\begin{itemize}
\item a vector whose entries are $cr_{i,c,t}$, the numbers of crimes of types $t\in \TT$ observed in $c$ in month $i$, where $|\TT| = 34$ according to Illinois Uniform Crime Reporting (IUCR) Codes \cite{dataset};
\item same vector with entries $cr_{i,a,t}$ for community $a\in \CC$;
\item same vector with entries $cr_{i,b,t}$ for community $b\in \CC$;
\item $(p_{i,c}, p_{i,a}, p_{i,b})$ - numbers of police stations connected to $a, b$ and $c$;
\item $(l_{i,c}, l_{i,a}, l_{i,b})$ - numbers of visitors to libraries connected to $a, b$ and $c$ in month $i$;
\item $(s_{i,c}, s_{i,a}, s_{i,b})$ - numbers of schools connected to $a, b$ and $c$ in month $i$;
\item $(sc_{i,c}, sc_{i,a}, sc_{i,b})$ - numbers of 311 service calls registered in $a, b$ and $c$ in month $i$.
\end{itemize}
The resulting vectors $p_{i,c}$ were corrected using min-max normalization and used as an input for prediction models. 

\subsection{Crime prediction}\label{sub:prediction}
To predict the number of crimes in a given community for a given year and month, we apply and compare three types of regression models, namely, polynomial regression,  support vector regression, and auto-regressive model. Each technique is briefly explained below. 

\subsubsection{Polynomial Regression}

Initially, we tried to use linear regression but because the obtained results were not satisfactory, we shifted to polynomial regression of degree two \cite{safro:highorder}. Experiments with  degree greater than two indicate that this method suffers from over-fitting the data. A polynomial regression of degree two is a process of  fitting a quadratic polynomial to the data. In our problem, we do not anticipate high-oscillatory shape of the data because under very rare circumstances the number of crimes in the megalopolis can quickly change in any direction. However, it is typically also not a straight line (like in linear regression) because the number of crimes exhibit periodic behavior. If $x_o, x_1, ..., x_n$ represent our feature set and $y$ represents the number of crimes of a certain type then the  relationship between $y$ and feature set can be expressed as 
\begin{equation}\label{eq:poly_reg}
y = a_0 + a_1x_1 + a_2x_1^2 + a_3x_2 + a_4x_2^2 + ... + a_{n-1}x_{n} + a_nx_{n}^2
\end{equation}
where $a_0, a_1, ..., a_n$ are the coefficients of features $x_i$ which are also called weights. By using either pseudo-inverse or gradient descent, weights can be determined. 

\subsubsection{Support Vector Regression}
The support vector regression (SVR) technique \cite{drucker1997support} uses same principles as the famous classification algorithm support vector machine (SVM) \cite{Vapnik:95}. In the core of SVM there is a convex quadratic programming problem (QP) in which support vectors that identify a separating hyperplane from the rest of the training data are discovered. In the support vector regression, the goal is to predict the data point value rather than its discrete class. In our implementation, we used LibSVM \cite{chang2011libsvm} to predict the number of crimes of a certain type for a given community and RBF function was chosen as the kernel. For our size of the data, the complexity of the kernel-based SVR was not prohibitive, however, we observed that in some cases, a tuning of regularization and kernel parameters with model selection methods is beneficial. For this purpose and for a larger data there exist several scalable techniques such as \cite{safro:engsvm,you2015svm}.

\subsubsection{Auto-regressive model}
In \cite{role}, the auto-regressive model was suggested for the analysis of crime data. Similar to other linear models, the variable of interest in auto-regressive model is predicted using a linear combination of its previous values. However, this model also adds a stochastic term on which the variables depend which is attractive for crime prediction because it makes an attempt to model data uncertainty that we have in crime prediction which in reality depends on many social factors. 
Because we just had 5 years of data, we chose the lag in this model to 2. That is, previous two year's crime values becomes the input for the prediction model. 

\begin{figure*}[t]
	\includegraphics[width=0.8\textwidth]{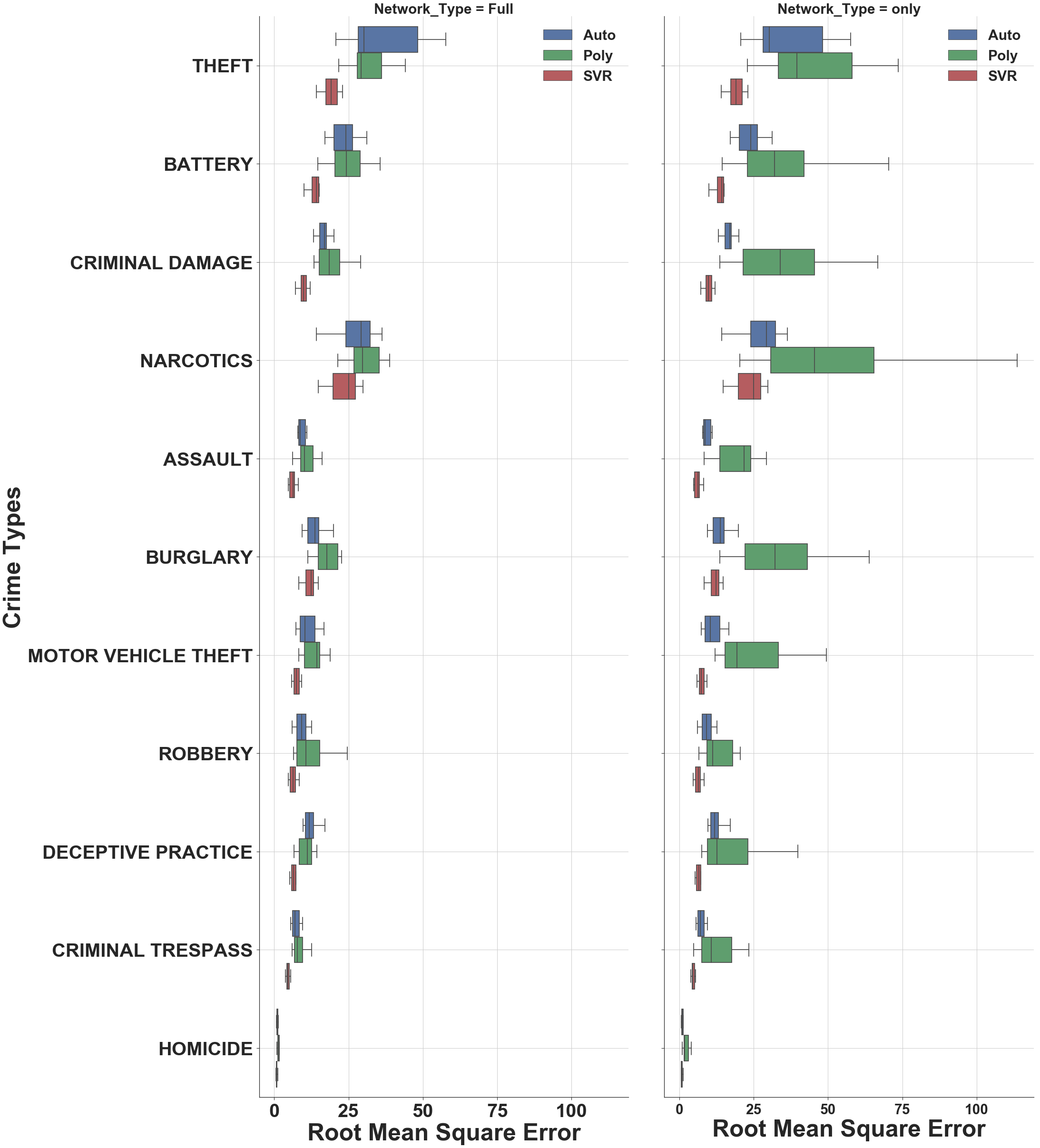}
	\caption{Comparison of \emph{full} and \emph{only crime} network based predictions of all three methods. Boxes colored blue, green, and red represent auto-, polynomial, and support vector regression methods, respectively. Only top 11 (i.e., most frequent) crime types are presented in this comparison. Each box represents the results of RMSE prediction over all 12 months of the year in all communities in the City of Chicago. The left and right parts of the figure correspond to \emph{full} and \emph{only crime} network based predictions.}
	\label{fig:fig2}
\end{figure*}

\begin{figure*}
	\includegraphics[width=0.75\textwidth,height=0.85\textheight]{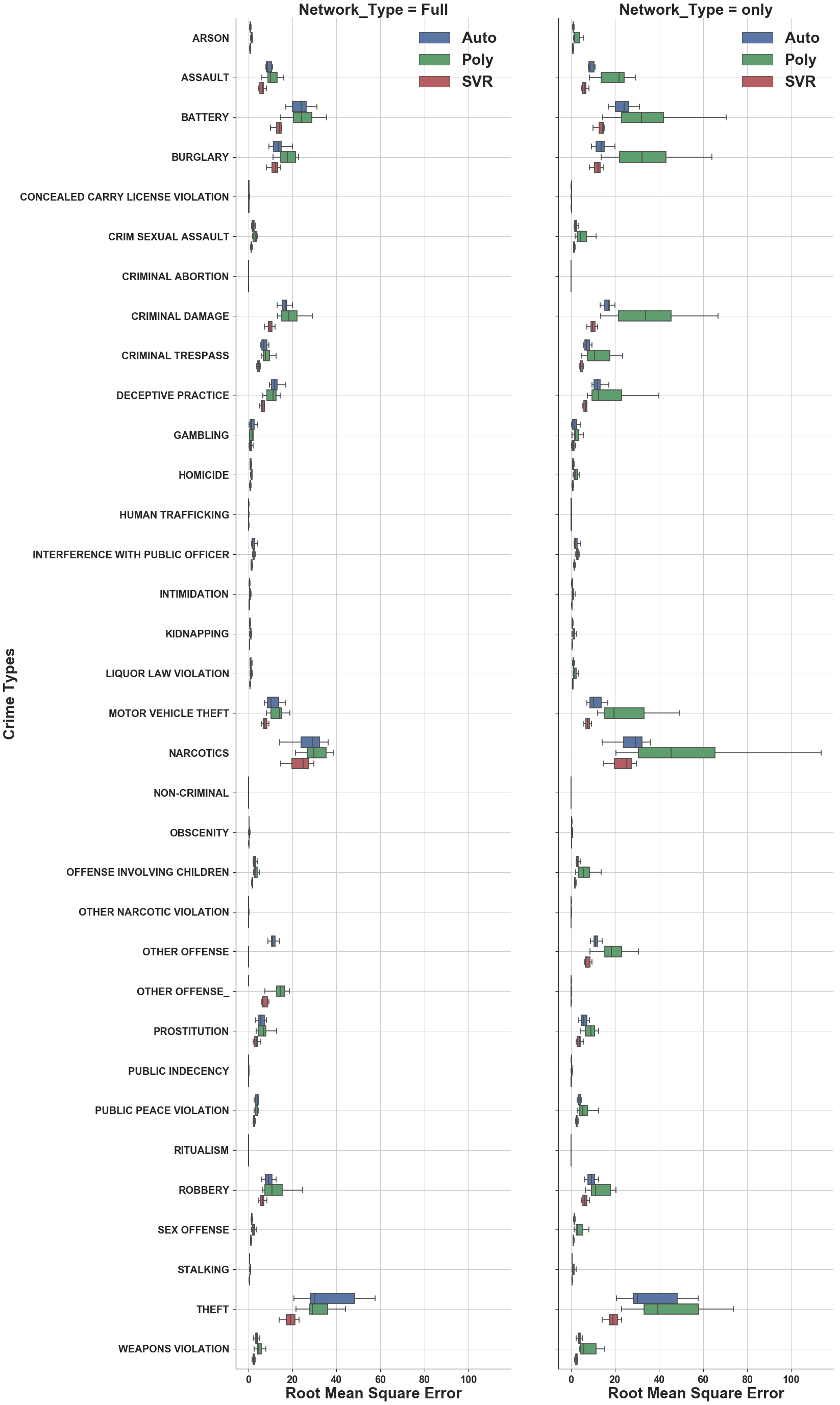}
	\caption{Comparison of \emph{full} and \emph{only crime} network based predictions of all three methods. Boxes colored blue, green, and red represent auto-, polynomial, and support vector regression methods, respectively. All crime types are presented in this comparison. Each box represents the results of RMSE prediction over all 12 months of the year in all communities in the City of Chicago. The left and right parts of the figure correspond to \emph{full} and \emph{only crime} network based predictions.}
	\label{fig:fig3}
\end{figure*}

\section{Computational Results}\label{results}
\begin{table}[h]
\begin{tabular}{|p{0.15\textwidth}|llll|l|}
\hline
{\bf Crime type}                         & {\bf 2011}  & {\bf 2012}  &{\bf  2013}  & {\bf 2014}  & {\bf 2015}    \\
\hline
Arson                             & 504   & 469   & 364   & 397   & 453   \\
Assault                           & 20411 & 19898 & 17971 & 16900 & 17041 \\
Battery                           & 60458 & 59134 & 54003 & 49447 & 48910 \\
Burglary                          & 26619 & 22844 & 17894 & 14570 & 13183 \\
Concealed carry license violation &       &       &       & 15    & 34   \\
Crime sexual assault              & 1471  & 1409  & 1272  & 1325  & 1365  \\
Criminal damage                   & 37332 & 35854 & 30853 & 27798 & 28672 \\
Criminal trespass                 & 8659  & 8215  & 8135  & 7539  & 6401  \\
Deceptive practice                & 12569 & 13515 & 13581 & 15466 & 15676 \\
Gambling                          & 736   & 724   & 596   & 393   & 310   \\
Homicide                          & 437   & 505   & 422   & 426   & 499   \\
Human trafficking                 &       &       & 2     & 2     & 13    \\
Interference with public officer  & 1048  & 1228  & 1281  & 1398  & 1308  \\
Intimidation                      & 171   & 156   & 134   & 116   & 122   \\
Kidnapping                        & 266   & 236   & 242   & 220   & 190   \\
Liquor law violation              & 619   & 573   & 465   & 397   & 292   \\
Motor vehicle theft               & 19387 & 16492 & 12582 & 9912  & 10070 \\
Narcotics                         & 38605 & 35488 & 34127 & 28995 & 23837 \\
Non-criminal                      &       & 6     & 7     & 27    & 35    \\
Non-criminal (subject specified)  &       & 2     &       & 1     &         \\
Obscenity                         & 40    & 26    & 24    & 38    & 46    \\
Offense involving children        & 2329  & 2197  & 2331  & 2358  & 2265  \\
Other narcotic violation          & 5     & 6     & 5     & 10    & 5     \\
Other offense                     & 20189 & 17479 & 17988 & 16972 & 17541 \\
Prostitution                      & 2424  & 2204  & 1652  & 1626  & 1322  \\
Public indecency                  & 13    & 17    & 10    & 10    & 14    \\
Public peace violation            & 3095  & 3007  & 3135  & 2903  & 2422  \\
Robbery                           & 13982 & 13485 & 11820 & 9800  & 9638  \\
Sex offense                       & 1071  & 1051  & 1019  & 958   & 972   \\
Stalking                          & 181   & 207   & 153   & 140   & 154   \\
Theft                             & 75146 & 75458 & 71524 & 61548 & 57319 \\
Weapons violation                 & 3880  & 3907  & 3246  & 3114  & 3362 \\
\hline
\end{tabular}
\caption{Total number of crimes  of all types registered in the City of Chicago Police Department in training (2011-2014) and test (2015) sets.}\label{tab:allcrimes}
\end{table}

The goal of computational experiments was to demonstrate a high-quality prediction of the number of crimes for each crime type in each community within the city of Chicago. This significantly extends experiments presented in \cite{role} and introduces new research directions. Because predicting the crimes is a quite complicated task because of the dependence of data on a variety of social, economic and political factors, our second goal was to demonstrate that network constructed and used to fuse various data sources does improve the quality of prediction. To the best of our knowledge, no work demonstrates a prediction of all crime types for each community on this and other comparable datasets, so, unfortunately no fair comparison with other methods can be presented. Our code and results are available at \url{https://goo.gl/V1yVLk}.

The comparison is presented in Figures \ref{fig:fig2}, and \ref{fig:fig3} using box plots of all three techniques and all crime types for 12 months in 2015. The training set was chosen to be 2011-2014. In Table \ref{tab:allcrimes}, we present the number of registered crimes in both sets. The quality of prediction is evaluated using Root Mean Square Error (RMSE) over all communities for a certain crime type $t\in \TT$ (i.e., the more the result or mean value is closer to zero the more accurate is the prediction), namely,
\[
\textsf{RMSE} (t) = \sqrt{\frac{\sum_{m=1}^{12} \sum_{c\in \CC} (cr_{i,c,t} - \tilde{cr}_{i,c,t})^2 }{\textsf{total \# of crimes of type } t}},
\]
where $\tilde{cr}_{i,c,t}$ is the predicted number of crimes of type $t$ in month $i$, at community $c$. For each crime type, the corresponding RMSE box was computed over 77 (communities) $\times$ 12 (months in test year 2015) = 924 values.

In order to demonstrate how other social information can affect the prediction we have created two types of networks which help us to identify two most similar communities, namely, \emph{only-crime} and \emph{full} networks. The only-crime network's set of nodes contains only three layers: communities, crimes, and police stations, $V=\CC\cup \TT \cup \PP$, i.e., the information that can be extracted from the police department dataset. The edge set generation and the rest of network construction remained the same. In the full network, we fuse information from all datasets as we describe in Section \ref{experiment}. 

In Figure \ref{fig:fig2}, we show 11 most frequent types of crimes that the police identifies in megalopolises. The usefulness of full network and data fusion in the prediction is immediately observed when the polynomial regression (green boxes) is applied. All box indicators, namely, median, maximum, minimum, first and third quartiles indicate an improvement of the prediction quality. The support vector regression (red boxes) always demonstrate the highest quality prediction that in many cases significantly improves the auto-regression proposed in \cite{role}. 

In Figure \ref{fig:fig3}, we present a similar comparison for all 34 types of crimes and observe similar results with SVR being a clear winner out of three regression methods. An important conclusion  we make based on our results is as follows. The SVR method (similar to regular support vector machines) maximizes the separating margin using support vectors discovered in training data. Usually, the number of support vectors is incomparably smaller than the training set size, and in our case, we observe that the number of support vectors is very small for all crime types and most communities. Because the number of support vectors is small and the prediction quality is high, their analysis can be \emph{fundamentally interpreted} by the domain experts and decision makers which can be used is applicable and novel crime preventing policies.   

\section{Future Work and Conclusion}\label{future}

We presented a prediction method for spatio-temporal information on crimes based on publicly available city data. We fuse the crime records with information about schools, libraries, police stations, and 311 service calls using network analytic approach to identify observations that improve the quality of prediction. After all the features of data are identified, regression-based prediction can be applied. We experimented with polynomial, auto-regressive and support vector regression methods and found that the quality of support vector regression significantly outperforms other methods. During our experiments, we have analyzed the Chicago crime events of all available 34 types and predicted them for all 12 months of 2015 for all 77 communities that divide the City of Chicago.

Although the data sources we proposed to work with are updated on-line (or at least there is always an option for the federal authorities to receive this information on-line), predicting crimes may benefit from features added from social networks as it has been shown in \cite{aghababaei2016mining,wang2012spatio,gerber2014predicting}. One of the promising research direction is fusing social network data that can provide socio-behavior "signals" for crime prediction. While the hypothesis that publicly available data in social networks may include predictive variables for crime and security events  \cite{aghababaei2016mining,avudaiappan2017detecting} is known, we argue that adding community-based information will improve the prediction quality.




\bibliographystyle{plain}
\bibliography{sample}

\end{document}